# Multimodal CNN Networks for Brain Tumor Segmentation in MRI: A BraTS 2022 Challenge Solution

Ramy A. Zeineldin [1,2,3], Mohamed E. Karar[2], Oliver Burgert[1], Franziska Mathis-Ullrich[3]

[1] Research Group Computer Assisted Medicine (CaMed), Reutlingen University, Germany
[2] Faculty of Electronic Engineering (FEE), Menoufia University, Egypt
[3] Health Robotics and Automation (HERA), Karlsruhe Institute of Technology, Germany
`Ramy.Zeineldin@Reutlingen-University.DE`

**Abstract.** Automatic segmentation is essential for the brain tumor diagnosis, disease prognosis, and follow-up therapy of patients with gliomas. Still, accurate detection of gliomas and their sub-regions in multimodal MRI is very challenging due to the variety of scanners and imaging protocols. Over the last years, the BraTS Challenge has provided a large number of multi-institutional MRI scans as a benchmark for glioma segmentation algorithms. This paper describes our contribution to the BraTS 2022 Continuous Evaluation challenge. We propose a new ensemble of multiple deep learning frameworks namely, DeepSeg, nnU-Net, and DeepSCAN for automatic glioma boundaries detection in pre-operative MRI. It is worth noting that our ensemble models took first place in the final evaluation on the BraTS testing dataset with Dice scores of 0.9294, 0.8788, and 0.8803, and Hausdorff distance of 5.23, 13.54, and 12.05, for the whole tumor, tumor core, and enhancing tumor, respectively. Furthermore, the proposed ensemble method ranked first in the final ranking on another unseen test dataset, namely Sub-Saharan Africa dataset, achieving mean Dice scores of 0.9737, 0.9593, and 0.9022, and HD95 of 2.66, 1.72, 3.32 for the whole tumor, tumor core, and enhancing tumor, respectively. The docker image for the winning submission is publicly available at (https://hub.docker.com/r/razeineldin/camed22).

## 1 Introduction

Glioma is the most common tumor type of tumor to originate in the brain and arises from the supportive tissue of the brain, called glial cells. Diagnosis of glioma is often challenging because of its invasive nature, extreme heterogeneity, its ability to occur in any part of the brain, and its sub-regions of varying shapes and sizes including the enhancing tumor (ET), peritumoral edema (ED), and the necrotic and non-enhancing tumor core (NET) [1, 2].

Manual segmentation of gliomas is the process of manually identifying and outlining the extent of glioma on medical imaging scans, such as magnetic resonance imaging (MRI) scans [3]. This is typically done by a trained medical professional, for example, a radiologist. One of the main challenges of manual glioma segmentation is its time-consuming and labor-intensive nature, requiring the person doing the segmentation to carefully review the images and outline the tumor. Additionally, manual segmentation can be subject to human error, causing inaccuracies in the final segmentation. Consequently, it is difficult to accurately assess the size and location of the tumor, which may have a negative impact on the treatment planning procedures.

The Brain Tumor Segmentation (BraTS) Challenge is an annual competition organized by the Medical Image Computing and Computer-Assisted Interventions (MICCAI) [4, 5]. The BraTS challenge is designed to encourage research in the field of medical image segmentation, with a focus on segmenting brain tumors in MRI scans. The challenge participants are provided with a dataset of MRI scans and asked to develop automated algorithms that can accurately segment the tumors in the images. BraTS 2022 Continuous Evaluation (BraTSCE) utilizes the largest annotated and publicly available multi-parametric (mpMRI) dataset provided by the BraTS 2021 challenge [2, 6, 7] as a common benchmark to foster the development of algorithms that can assist doctors in the diagnosis and treatment of brain tumors.

Deep learning is a sub-field of artificial intelligence that uses neural networks to learn from data and make predictions. In the context of glioma segmentation, deep learning algorithms can be trained to analyze MRI scans of the brain and identify areas that contain tumors. More specifically, the encoder-decoder architecture with skip connections, first introduced by the U-Net [8, 9], has gained popularity in the medical field outperforming other traditional methods in brain glioma segmentation [10-12]. The use of deep learning in glioma segmentation can help to automate the process, making it faster and more efficient. By using deep learning, researchers and doctors can more quickly and accurately identify tumors in MRI scans, which can ultimately improve patient care. In the context of the BraTS challenge, the recent winning contributions of 2019 [13], 2020 [14], and 2021 [11] extend the U-Net architecture by adding two-stage cascaded U-Net [13], making significant architecture changes [14], or using ensemble predictions.

In this paper, we extend our previous work [12] by proposing a fully automated convolutional neural network (CNN) method for glioma segmentation based on an ensemble of three encoder-decoder methods, namely, DeepSeg [15], our earlier deep learning framework for automatic brain tumor segmentation using two-dimensional T2 Fluid Attenuated Inversion Recovery (FLAIR) scans, nnU-Net [14], a self-configuring method for automatic biomedical segmentation, and DeepSCAN [16] architecture which contains densely connected blocks of dilated convolutions. The remainder of the paper is organized as follows: The BraTS dataset and the encoder-decoder CNN architecture are described in Section 2. Section 3 presents the experimental methods of the ensemble model, whereas the paper is concluded in Section 4.

## 2 Materials and Methods

### 2.1 Dataset

The MRI volumes have been used for the training and evaluation of models based on the Multi-modal BraTS Challenge 2021 [2]. The BraTS 2021 training dataset contains 1251 scans along with truth annotations of tumorous regions. For each case, BraTS provide four modalities: native (T1), post-contrast T1-weighted (T1Gd), T2-weighted (T2), and FLAIR. Ground truth segmentation consists of four classes: enhancing tumor (ET) (label 4), peritumoral edema (ED) (label 2), necrotic tumor core (NCR) (label 1), and background (label 0). These sub-regions can be clustered together to compose three semantically meaningful tumor classes enhancing tumor (ET), the addition of ET and NCR represents the tumor Core (TC) region, and the addition of ED to TC represents the whole tumor (WT). The BraTS 2021 database includes also 219 cases that were used for the public leaderboard during the validation phase, in addition to 530 cases for the final ranking of the participants.

Initial preprocessing steps were performed by the BraTS including co-registration to the same anatomical template, isotropic resampling to $1\text{mm}^3$ resolution, and skull-stripping. The resultant MRI volumes and associated labels are of shape $240 \times 240 \times 155$. The provided data were further processed prior to being fed into the networks. The MRI volumes were cropped to non-zero voxels to reduce the computation with a spatial resolution of $192 \times 224 \times 160$. Since the intensity in MR images is qualitative, the voxels were normalized by their mean and standard deviation for each input MRI image.

### 2.2 Neural Network Architectures

**DeepSeg.** Figure 1 outlines our previously proposed model [12, 15], which is a modular CNN framework for fully automatic brain tumor detection and segmentation. Inspired by the U-Net, DeepSeg consists of a contracting path followed by an expansive path. The contracting path is made up of a series of $3 \times 3 \times 3$ convolutional and $2 \times 2 \times 2$ max-pooling layers, which extract hierarchical features from the input image. The expansive path, on the other hand, is made up of a series of $2 \times 2 \times 2$ deconvolutional layers, which use those hierarchical features to upsample the output of the contracting path and produce a segmentation map for the input image. DeepSeg utilizes recent advances in CNNs including dropout, batch normalization (BN), and rectified linear unit (ReLU) [17, 18]. The initial filter size of convolutional kernels is set to 8 and doubled at the following layers which allow the network to learn features at multiple scales and improve its performance on the segmentation task. Finally, a $1 \times 1 \times 1$ convolutional layer followed by a softmax function is employed for the output segmentation.

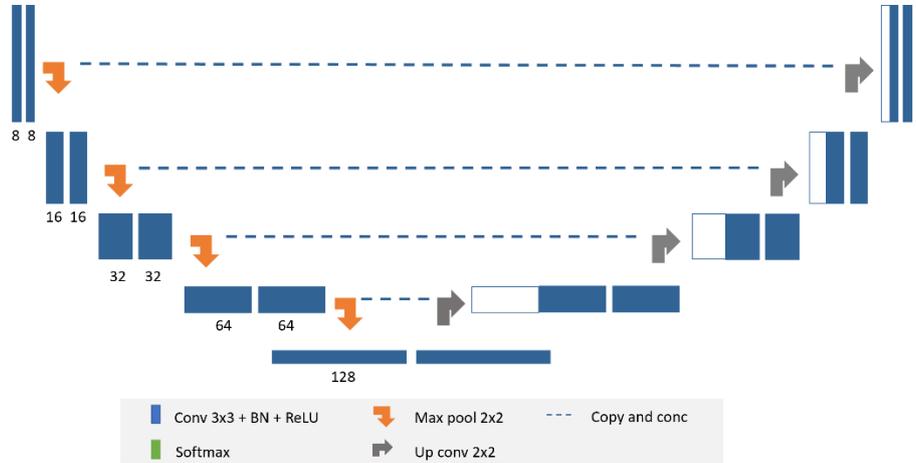

**Fig. 1.** DeepSeg architecture, as applied to brain tumor segmentation in BraTS 2021 challenge [12].

**nnU-Net** is a deep-learning framework for medical image segmentation [10]. It is an extension of the U-Net architecture, which is a popular deep-learning architecture for image segmentation tasks. In contrast to DeepSeg, nnU-Net does not employ any of the recently proposed architectural advances in deep learning and is based only on plain convolutions for feature extraction. More specifically, nnU-Net uses strided convolutions for downsampling whereas transposed convolutions are applied for upsampling. Figure 2 outlines the enhanced nnU-Net incorporating three main modifications that led to the first rank in the segmentation task of the BraTS challenge in 2021 [11]. First, the network size is asymmetrically increased by doubling the number of filters in the encoder while maintaining the same filters in the decoder. Second, group normalization layers are replaced by batch normalization which has been shown to work better for the low batch size. Third, the employment of a self-attention mechanism or transformer [19] in the decoder allows the model to focus on different parts of the input image at different times, which can be useful for identifying and segmenting complex glioma sub-structures.

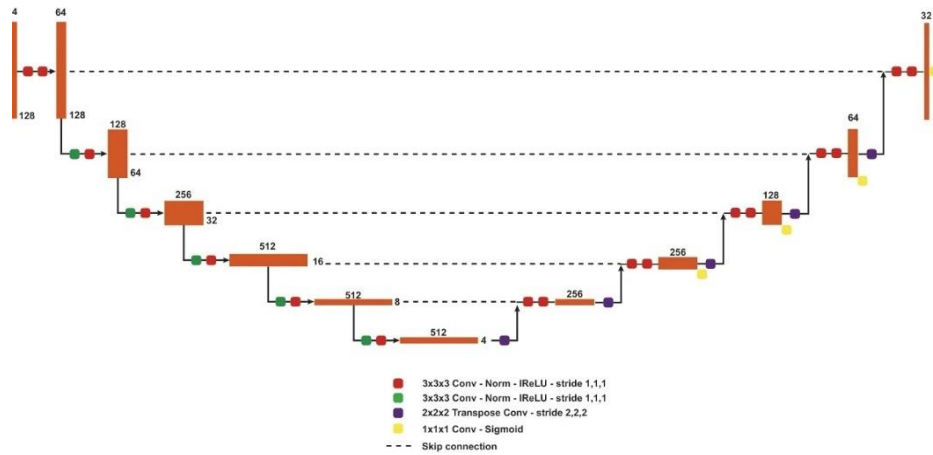

**3x3x3 Conv - Norm - IReLU - stride 1,1,1**
**3x3x3 Conv - Norm - IReLU - stride 1,1,1**
**2x2x2 Transpose Conv - stride 2,2,2**
**1x1x1 Conv - Sigmoid**
**- - - - Skip connection**

**Fig. 2.** Enhanced nnU-Net network, as applied to brain tumor segmentation in BraTS 2021 challenge [11].

**DeepSCAN.** Figure 3 shows the two DeepSCAN architectures introduced for brain tumor segmentation. Inspired by the recent Densenet architecture [20] and U-Net [8], DeepSCAN architecture was proposed for semantic segmentation. Instead of using transition layers and pooling operations, dilated convolutions are used to increase the receptive field of the encoder. Similar to Densenet, the output of each layer is concatenated with its input before passing to the next layer. Moreover, label uncertainty is applied directly to the loss function, which allows the prediction of the CNN to be involved in evaluating the network decision. This hybrid 2D/ 3D approach led to more stable results and ranked third in the BraTS 2018 challenge.

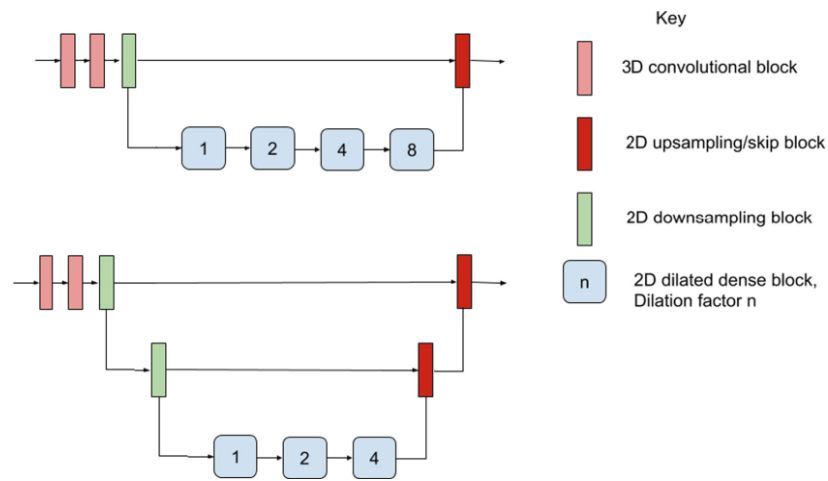

**Key**

**3D convolutional block**
**2D upsampling/skip block**
**2D downsampling block**
**n  2D dilated dense block, Dilation factor n**

**Fig. 3.** DeepSCAN architectures, as applied to brain tumor segmentation in BraTS 2018 challenge [16].

### 2.3 Ensemble Learning

The ensemble is a technique in machine learning where multiple models are trained and combined to make a single prediction [21, 22]. In the context of glioma segmentation, an ensemble of models could be used to improve the performance of the segmentation process and provide more accurate results. For example, multiple models could be trained on different subsets of the data, and their predictions could be combined to create a more accurate segmentation of the tumor in an MRI scan.

In this paper, we used three different CNN models, namely, DeepSeg [15], nnU-Net [10], and DeepSCAN [16] which follow the U-Net pattern [8, 9] and consist of encoder-decoder architecture interconnected by skip connections (as discussed in Section 2.3). The final results were obtained by using the Simultaneous Truth and Performance Level Estimation (STAPLE) [23], which is an expectation-maximization ensemble method used in medical image segmentation. The STAPLE method works by estimating the truth, or ground truth, and the performance level of each model in the ensemble. It then combines the predictions of the models to create a final segmentation that is more accurate than any of the individual models.

### 2.4 Post-processing

Post-processing is a step in the glioma segmentation process that follows the initial segmentation of the tumor. It typically involves refining the initial segmentation to improve its accuracy and reduce any errors or inconsistencies. These methods could include techniques such as morphological operations, region growing, or level set evolution. The goal of post-processing is to produce a final segmentation that is as accurate and reliable as possible.

In the BraTS Challenge, the segmentation of the tumor core, and determining the small blood vessels (necrosis or edema), is particularly challenging. This is especially apparent in low-grade glioma (LGG) patients where there may be no enhancing tumor and, therefore, the BraTS challenge evaluates the segmentation as binary values of 0 or 1. Nevertheless, a Dice value of 0 will be generated for the scenario in which there are only small false positives in the predicted segmentation map of a patient with no enhancing tumor. To overcome this problem, all enhanced tumor outputs were re-labeled with necrotic (label 1) when the total predicted ET region is lower than a threshold although this strategy may have the side effect of eliminating some correct predictions.

## 3 Experiments and Results

### 3.1 Cross-validation Training

DeepSeg model was implemented in Tensorflow [25] while nnU-Net and DeepSCAN were implemented in PyTorch [26] frameworks. Pre-trained models were used for DeepSCAN and nnU-Net to reduce the training time, ensuring the same results as in their previous BraTS challenge contributions [11, 16]. For training DeepSeg, five-fold cross-validation was used for training each model on the 1251 cases of the BraTS 2021 training dataset for a maximum of 1200 epochs. This allows each model to be trained and evaluated multiple times using different combinations of training and testing data, which can provide a more accurate estimate of the performance of a model. Adam optimizer [24] has been applied with an initial learning rate of $1e^{-4}$ and a default value of $1e^{-7}$ for epsilon. All experiments were conducted on a single Nvidia GPU (RTX 2080 Ti GPU with 11 GB VRAM or RTX 3060 with 12 GB VRAM). Randomly sampled patches of $128 \times 128 \times 128$ voxels are input to our networks with batch sizes varying from 2 to 5 and the post-processing threshold is set to 200 voxels. This tiling strategy allows the model to be trained on multi-modal high-resolution MRI images with low GPU memory requirements. To overcome the effect of class imbalance between tumor labels and the brain healthy tissue, on-the-fly spatial data augmentations during training have been applied including random rotation between 0 and $30°$, random 3D flipping, power-law gamma intensity transformation, or a combination of them.

### 3.2 Online Evaluation

Table 1 summarizes the results of the models on the BraTS 2021 validation, where the five-fold cross-validation for each model is averaged as an ensemble. Two evaluation metrics are used for the BraTS 2021 benchmark, computed by the online evaluation platform of Sage Bionetworks Synapse (Synapse)[1], which are the DSC and the Hausdorff distance (95%) (HD95). Similarly, we computed the averages of DSC scores and HD95 values across the three evaluated tumor sub-regions and then used them to rank our methods in the final column.

By using a region-based version of DeepSeg with an input patch size of $128 \times 128 \times 128$ voxels, batch size of 5, applied post-processing stage, and on-the-fly data augmentation, the DeepSeg model achieved good results of DSC values of 0.8356, 0.8508, and 0.9137 for the ET, TC, and WT regions, respectively. Additionally, we used two different models of nnU-Net [14], the BraTS 2020 winning approach, and DeepSCAN, one of the BraTS 2018 winning approaches. The first model, nnU-Net, is a region-based version of the standard nnU-Net, large batch size of 5, more aggressive data augmentation as described in [14], trained using batch Dice loss, and including the postprocessing stage. DeepSCAN model is similar to nnU-Net and DeepSeg model with the output layer as three logits, one for the whole tumor, tumor core, and enhancing regions rather than using a softmax layer. DeepSCAN model ranks third in our ranking (see Table 1) achieving the best average HD95 of 8.7886 while maintaining a good DSC of 0.8739.

For the BraTSCE 2022 challenge, we selected the three top-performing models to build our final ensemble: DeepSeg + nnU-Net + DeepSCAN. It is worth mentioning that our final ensemble was implemented by first predicting the validation cases individually with each model configuration, followed by averaging the softmax outputs to obtain the final cross-validation predictions. After that, the STAPLE [23] was applied

---

to aggregate the segmentation produced by each of the individual methods using the probabilistic estimate of the true segmentation. This led to our best score of 0.8821 and 9.5440 for the mean DSC and HD95, respectively on the BraTS 2021 final validation dataset.

**Table 1.** Results of our five-fold cross-validation models on BraTS 2021 validation cases. All reported values were computed by the online evaluation platform Synapse. The average of DSC and HD95 scores are computed and used for ranking our methods.

| Model | DSC ↑ | | | | HD95 ↓ | | | | Rank |
|-------|-------|------|------|------|--------|------|------|------|------|
|       | ET    | TC   | WT   | Avg  | ET     | TC   | WT   | Avg  |      |
| DeepSeg | 0.8356 | 0.8508 | 0.9137 | 0.8667 | 17.75 | 11.56 | 4.15 | 11.15 | 4 |
| nnU-Net | 0.8402 | 0.8718 | 0.9213 | 0.8778 | 16.03 | 8.95 | 3.82 | 9.60 | 2 |
| DeepSCAN | 0.8306 | 0.8683 | 0.9228 | 0.8739 | **14.50** | 7.91 | 3.95 | **8.79** | 3 |
| Ensemble | **0.8438** | **0.8753** | **0.9271** | **0.8821** | 17.50 | **7.53** | **3.60** | 9.54 | 1 |

- Bold values correspond to higher scores

Table 2, Table 3, and Table 4 provide complete statistics of the ensemble model performance on the test datasets. All reported values were provided by the challenge organizers. Our method took first place in the BraTSCE 2022 competition on both the unseen BraTS and Sub-Saharan Africa (SSA) datasets[2]. However, the results were not so great on the third pediatric test dataset. This is particularly due to the inherent variability and complexity of the tumor tissue in pediatric data which makes segmentation of pediatric brain tumors challenging. Unlike other types of tumors, which tend to be more homogeneous, pediatric brain tumors can be highly heterogeneous, with different types of tissue and abnormal cells. Additionally, the small size and delicate nature of the brain tissue in pediatric patients can make it difficult to accurately identify and segment the tumor tissue. Other challenges in the segmentation of pediatric brain tumors include the need for specialized techniques, as well as the lack of large, annotated datasets for training and validation.

---

**Table 2.** Final ensemble results on the BraTS 2021 test dataset.

|  | DSC | | | HD95 | | |
|---|---|---|---|---|---|---|
|  | ET | TC | WT | ET | TC | WT |
| Mean | 0.8803 | 0.8788 | 0.9294 | 12.05 | 13.54 | 5.24 |
| StdDev | 0.1765 | 0.2384 | 0.1026 | 59.61 | 59.67 | 23.08 |
| Median | 0.9351 | 0.9637 | 0.9602 | 1.00 | 1.41 | 1.57 |
| 25quantile | 0.8633 | 0.9179 | 0.9210 | 1.00 | 1.00 | 1.00 |
| 75quantile | 0.9645 | 0.9814 | 0.9791 | 1.73 | 3.00 | 3.61 |

**Table 3.** Final ensemble results on the SSA test dataset.

|  | DSC | | | HD95 | | |
|---|---|---|---|---|---|---|
|  | ET | TC | WT | ET | TC | WT |
| Mean | 0.9022 | 0.9593 | 0.9737 | 3.32 | 1.72 | 2.66 |
| StdDev | 0.1303 | 0.0704 | 0.0468 | 6.09 | 1.44 | 6.50 |
| Median | 0.9492 | 0.9841 | 0.9872 | 1.41 | 1.00 | 1.00 |
| 25quantile | 0.9014 | 0.9725 | 0.9735 | 1.00 | 1.00 | 1.00 |
| 75quantile | 0.9608 | 0.9893 | 0.9904 | 3.00 | 1.41 | 1.00 |

**Table 4.** Final ensemble results on the pediatric test dataset.

|  | DSC | | HD95 | |
|---|---|---|---|---|
|  | TC | WT | TC | WT |
| Mean | 0.2639 | 0.7953 | 181.76 | 24.84 |
| StdDev | 0.3535 | 0.2465 | 179.51 | 80.43 |
| Median | 0.0150 | 0.8802 | 68.48 | 4.12 |
| 25quantile | 0.0000 | 0.8225 | 7.07 | 3.08 |
| 75quantile | 0.5948 | 0.9049 | 373.13 | 6.48 |

### 3.3 Qualitative Output

Figure 4 depicts the qualitative performance of the segmentation predictions. It shows results generated by the final ensemble model on the BraTS 2021 validation dataset. In the three rows, the best, median, and worse segmentations are shown according to their DSC scores. It is obviously that our proposed model achieved best results with overall high quality. However, applying our post-processing strategy showed a limitation as illustrated in Section 2.4. Nevertheless, the WT region was detected with a good quality (DSC of 0.9606) which could be valuable for future clinical use.

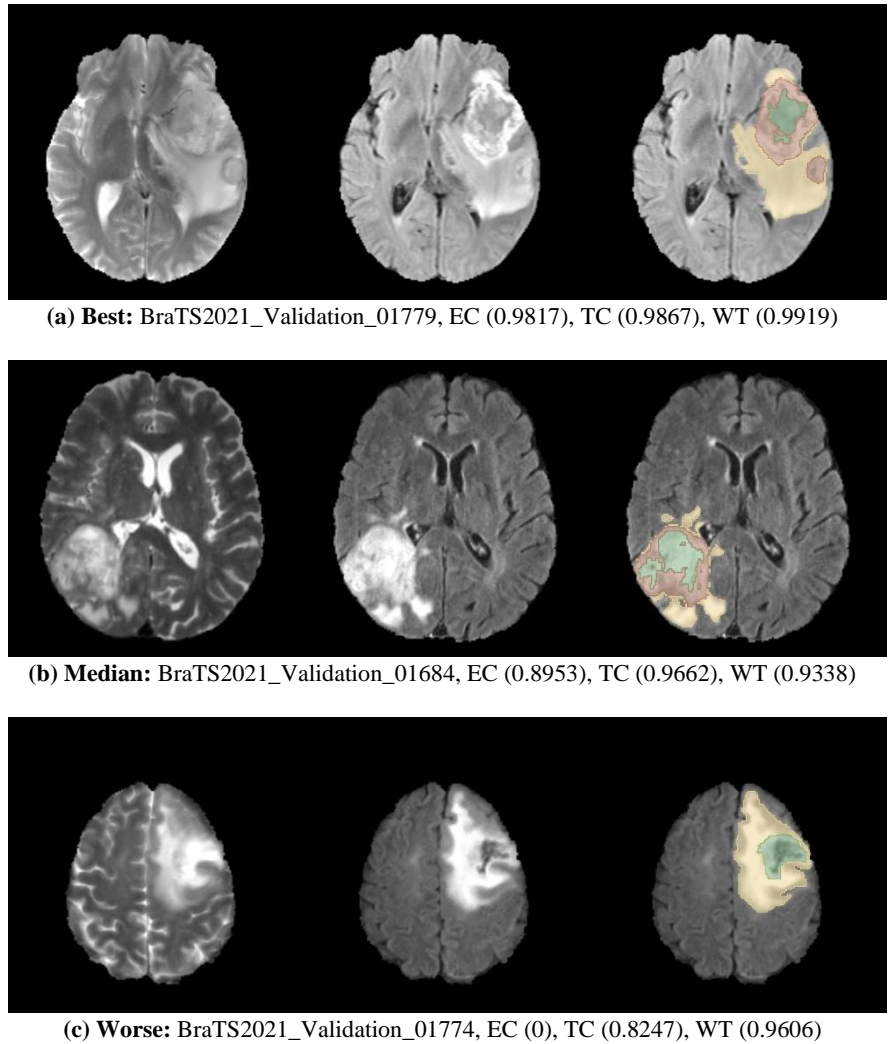

**(a) Best:** BraTS2021_Validation_01779, EC (0.9817), TC (0.9867), WT (0.9919)

**(b) Median:** BraTS2021_Validation_01684, EC (0.8953), TC (0.9662), WT (0.9338)

**(c) Worse:** BraTS2021_Validation_01774, EC (0), TC (0.8247), WT (0.9606)

**Fig. 4.** Sample qualitative validation set results of our ensemble model. The best, median and worse cases are shown in the rows. Columns display the T2, T2-FLAIR, and the overlay of our predicted segmentation on the T2-FLAIR image. WT includes all visible labels (green, yellow and red labels), TC is a union of green and red, while ET class is shown in green.

## 4 Conclusion

This paper presented our contribution to the segmentation task of the BraTSCE 2022 challenge. Ensemble models from multiple configurations of different state-of-the-art U-Net variants, namely, DeepSeg [12], nnU-Net [14], and DeepSCAN [16], have been explored. Based on our internal ranking strategy, the final submission was selected as the ensemble of these methods. Post-processing methods were used to improve the accuracy of the developed segmentation algorithms despite the side effects of ignoring some correct predictions. Table 1 lists the results of our methods on the validation set computed by the online evaluation platform Synapse. Remarkably, our method achieved DSC of 0.9271, 0.8753, and 0.8438 as well HD95 of 17.5041, 7.5326, and 3.5952 for, ET, TC, and WT regions on the validation dataset, respectively.

This ensemble method won the BraTSCE 2022 competition on two unseen test datasets. On the BraTS test dataset, our submission achieved DSC scores of 0.8803, 0.8788, and 0.9294 as well as HD95 of 12.05, 13.54, and 5.24 for ET, TC, and WT, respectively. Remarkably, our model obtained DSC scores of 0.9022, 0.9593, and 0.9737 as well as HD95 of 3.32, 1.72, and 2.66 for ET, TC, and WT on the SSA test dataset, respectively. In general, qualitative evaluation supports the numerical evaluation showing a high-quality segmentation. The findings suggest that this approach can be readily employed for clinical practice.

**Acknowledgments.** The first author is supported by the German Academic Exchange Service (DAAD) [scholarship number 91705803].